# Application of Radial Basis Network Model for HIV/AIDs Regimen Specifications

Dr. P.Balasubramanie, Kongu Engineering College, Perundurai.

Prof. M.Lilly Florence, Adhiyamaan College of Engineering, Hosur.

**Abstract:** HIV/AIDs Regimen specification one of many problems for which bioinformaticians have implemented and trained machine learning methods such as neural networks. Predicting HIV resistance would be much easier, but unfortunately we rarely have enough structural information available to train a neural network. To network model designed to predict how long the HIV patient can prolong his/her life time with certain regimen specification. To learn this model 300 patient's details have taken as a training set to train the network and 100 patients medical history has taken to test this model. This network model is trained using MAT lab implementation.

**Resources:** RBN algorithm implemented in MAT Lab, data from a ART centre in Tamil Nadu.

## 1. Introduction

The Human Immunodeficiency Virus is one of the main causes of death in the world. The HIV is a human pathogen that infects certain types of lymphocytes called T-helper cells, which are important to the immune system. Without a sufficient number of T-helper cells, the immune system is unable to defend the body against infections.

Clinical trail system for HIV/AIDs is a complex one. This is the case because every patient is unique with his/her own health, set of genetic traits, predisposition to side effects and prognosis. Additionally, many symptoms and diagnoses are inherently imprecise in their definition and difficult to measure. Although clinical trial data provide excellent information regarding excepted treatment outcomes for large groups of patients, the prediction of actual treatment outcomes and clinical courses for a particular individual patient may be subject to a considerable degree of uncertainty. In this paper, we will focus on the function of RBN for our problem and we have discussed about some of the outcomes of this trained network.

## 2. Radial Basis Function Network (RBFN) Model

The RBFN model consists of three layers, the input, hidden and output layer. The nodes within each layer are fully connected to the previous layer as shown in Fig 1. The input variables are each assigned to a node in the input layer and pass directly to the hidden layer without weights. The hidden nodes or units contain the radial basis functions also called transfer functions and are analogous to the sigmoid functions commonly used in the back propagation network models. They are represented by the bell shaped curve in the hidden model shown in figure1.

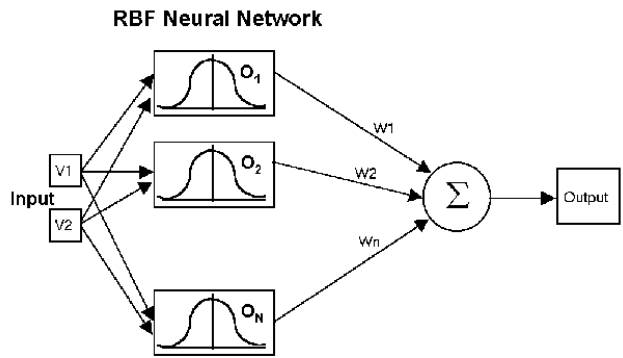

Figure 1  FBN Model for this network

1. Input layer – There is one neuron in the input layer for each predictor variable. In the case of categorical variables, N-1 neurons are used where N is the number of categories. The input neurons (or processing before the input layer) standardizes the range of the values by subtracting the median and dividing by the inter quartile range. The input neurons then feed the values to each of the neurons in the hidden layer.
2. Hidden layer – This layer has a variable number of neurons (the optimal number is determined by the training process). Each neuron consists of a radial basis function centered on a point with as many dimensions as there are predictor variables. The spread (radius) of the RBF function may be different for each dimension. The centers and spreads are determined by the training process. When presented with the x vector of input values from the input layer, a hidden neuron computes the Euclidean distance of the test case from the neuron's center point and then applies the RBF kernel function to this distance using the spread values. The resulting value is passed to the summation layer.
3. Summation layer – The value coming out of a neuron in the hidden layer is multiplied by a weight associated with the neuron (W1, W2, ...,Wn in this figure) and passed to the summation which adds up the weighted values and presents this sum as the output of the network. Not shown in this figure is a bias value of 1.0 that is multiplied by a weight W0 and fed into the summation layer. For classification problems, there is one output (and a separate set of weights and summation unit) for each target category. The value output for a category is the probability that the case being evaluated has that category.

## 3. Training RBF Networks

The following parameters are determined by the training process:

1. The number of neurons in the hidden layer.
2. The coordinates of the center of each hidden-layer RBF function.
3. The radius (spread) of each RBF function in each dimension.
4. The weights applied to the RBF function outputs as they are passed to the summation layer.

Various methods have been used to train RBF networks. One approach first uses K-means clustering to find cluster centers which are then used as the centers for the RBF functions. However, K-means clustering is a computationally intensive procedure, and it often does not generate the optimal number of centers. Another approach is to use a random subset of the training points as the centers.

The RBF network has a feed forward structure consisting of a single hidden layer of $J$ locally tuned units, which are fully interconnected to an output layer of $L$ linear units. All hidden units simultaneously receive the $n$-dimensional real valued input vector $X$ (Figure 3). The main difference from that of MLP is the absence of hidden-layer weights. The hidden-unit outputs are not calculated using the weighted-sum mechanism/sigmoid activation; rather each hidden-unit output $Z_j$ is obtained by closeness of the input $X$ to an $n$-dimensional parameter vector $m_j$ associated with the $j$th hidden unit[10,11]. The response characteristics of the $j$th hidden unit ( $j = 1, 2, ¼, J$) is assumed as,

$$Z_j = K(\|X - \mu_j\|) / \sigma^2_j$$

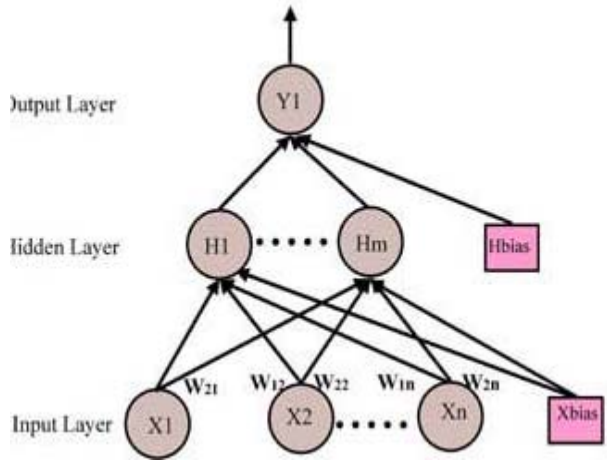

Figure 2. Feed forward neural network

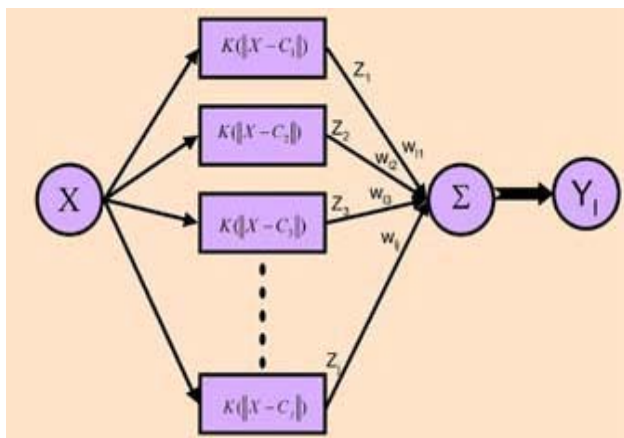

Figure 3 Radial basis function neural network

where $K$ is a strictly positive radically symmetric function (kernel) with a unique maximum at its 'centre' $mj$ and which drops off rapidly to zero away from the centre. The parameter $sj$ is the width of the receptive field in the input space from unit $j$. This implies that $Zj$ has an appreciable value only when the distance $\| \| j X - m$ is smaller than the width $sj$. Given an input vector $X$, the output of the RBF network is the $L$-dimensional activity vector $Y$, whose $l$th component ($l = 1, 2 ¼ L$) is given by,

l
$$Y_l(X) = \sum w_{lj} z_j(X)$$
j=1

For $l$ = 1, mapping of eq. (1) is similar to a polynomial threshold gate. However, in the RBF network, a choice is made to use radically symmetric kernels as 'hidden units'.

## 4. Experimental Setup and Result of RBN network Model

Consider an observation used to train the model to have r inputs variables such as patient's age, weight, CD count, HB rate, CD8 count and so on. The weight for calculating the sum is regimen specification. The regimen specification and other factors are not common for all the patients. According to the output received the weight is adjusted and again it executed until the required output is generated. Now the adjusted weight is the regimen specification for the particular patient. In this way that the network is trained.

In this research, we have taken 500 patients medical history . Among this, 300 cases used as training set and 200 cases used as testing set. Among these homogeneous set, the output of this model defines the regimen specification for two sets. One set of cases, they can prolong their life more than 10 years only if they should follow the specified regimens, another set, they are very difficult to prolong their life even if they follow the restricted specification of regimens. In this study the maximum age of the patients taken is 45 and only used homogenous data. Table 1 shows the details of patients (sample) . Table 2 shows the suggested regimen specification. The definition of regimen specification is consulted with a Physician in the famous ART centre in Tamil Nadu.

Table 1

| Patient ID | Age | Weight | CD4 | CD8 | HB | TLC | First Identified Date |
|---|---|---|---|---|---|---|---|
| A | 23 | 45 | 204 | 721 | 10 | 945 | 23.11.07 |
| B | 23 | 43 | 187 | 1498 | 10 | 1769 | 23.10.07 |
| C | 34 | 42 | 38 | 812 | 12.5 | 1076 | 28.07.06 |
| D | 35 | 35 | 28 | 940 | 8 | 1202 | 19.12.07 |

| E | 32 | 41 | 238 | 408 | 9 | 1100 | Year 2000 |
| F | 38 | 92 | 33 | 294 | 9.2 | 571 | 10.12.07 |
| G | 37 | 43 | 123 | 1262 | 8.5 | 1605 | 04.09.07 |
| H | 40 | 35 | 38 | 745 | 8 | 1169 | 01.12.07 |
| I | 42 | 43 | 34 | 512 | 12.5 | 811 | 23.09.98 |
| J | 39 | 40 | 112 | 643 | 10 | 900 | 869 |

**In Table 1** there 10 patients data has been taken, the important factors to define the regimen are given. The underlined data are conserved as set 2, since this case cannot prolong their life for the expected periods. The remaining data consider as set 1, they can prolong their life provided they should strictly follow the defined regimen specification.

Table 2

| Patient ID | Regimen | R_Specification | Prolong Period |
|---|---|---|---|
| A | ZLN | 2 Per Day | >75% |
| B | ZLN | 2 Per Day | >75% |
| C | ZLE | 2 Per Day | <50% |
| D | SLN 30 | 2 Per Day | <50% |
| E | ZLN | 2 Per Day | >75% |
| F | ZLE | 2 Per Day | <50% |
| G | SLN 30 | 2 Per Day | >75% |
| H | ZLN | 2 Per Day | <50% |
| I | ZLN | 2 Per Day | <50% |
| J | ZLN | 2 Per Day | >75% |

**In Table 2** the outcome of this RBN has discussed one set of cases they can prolong their life time more than 75% i.e in our research we have taken maximum 10 years, so more than 7.5 yrs they can alive if they give the continuous response. The second set of cases they can prolong maximum of 50% of the period. There are three combination of regimens would be prescribed based on their TLC count for them. The specification is depends on their weight, HB and CD4 count. The grams and mill grams of this regimens depends on their age and weight.

## 5. Conclusion

The sensitivity and specificity of both neural network models had a better predictive power compared to logistic regression. Even when compared on an external dataset, the neural network models performed better than the logistic regression. This study indicates the good predictive capabilities of RBF neural network. Also the time taken by RBF is less than that of MLP in our application. Though application of RBF network is limited in biomedicine, many comparative studies of MLP and statistical methods are used. The limitation of the RBF neural network is that it is more sensitive to dimensionality and has greater difficulties if the number of units is large. Generally, neural network results presented are mostly based only on the same dataset. Here an independent valuation is done using external validation data and both the neural network models performed well, with the RBF model having better prediction. The predicting capabilities of RBF neural network had showed good results and more applications would bring out the efficiency of this model over other models. ANN may be particularly useful when the primary goal is classification and is important when interactions or complex nonlinearities exists in the dataset. Logistic regression remains the clear choice when the primary goal of model development is to look for possible causal relationships between independent and dependent variables, and one wishes to easily understand the effect of predictor variables on the outcome. There have been ingenious modifications and restrictions to the neural network model to broaden its range of applications. The bottleneck networks for nonlinear principle components and networks with duplicated weights to mimic autoregressive models are recent examples. When classification is the goal, the neural network model will often deliver close to the best fit. The present work was a part of our research.